\documentstyle[version,12pt]{article}
\def\Message{\immediate\write16}
\def\inprod{\mathop{\kern -0.05em\raise -0.1em\hbox{%
  \vrule height 0.03em width 0.6em depth 0em%
  \vrule height 0.7em width 0.03em depth 0em}\kern 0.1em}\nolimits}
\def\d{\mbox{\sf d}}

\def\be{\begin{equation}}
\def\ee{\end{equation}}
\Message{}
\Message{If you have RUSSIAN TEX you can get Russian version}
\Message{of this article.}
\Message{For this purpose you should make minimal changes}
\Message{in the source file.}
\Message{Instructions are given at the beginning of it.}
\Message{}
\Message{Если в Вашем распоряжении есть русский ТеХ,}
\Message{Вы можете  получить русскую версию этой работы.}
\Message{Для этого Вам нужно сделать минимальные изменения}
\Message{в исходном файле.}
\Message{Инструкции даны в начале файла.}
\begin{document}
\excludeversion{Russian}
\includeversion{English}
\excludeversion{rprep}
\begin{English}
\title{On the difference between admissible and
``differentiable'' Hamiltonians}
\author{Vladimir O. Soloviev\\
\small Institute for High Energy Physics\\
\small  Protvino, 142284, Moscow Region, Russia\\
\small e-mail: vosoloviev@mx.ihep.su}
\date{October 30, 1996}
\maketitle
\begin{abstract}
It is shown that the Regge-Teitelboim criterion for fixing
the unique boundary contribution to the Hamiltonian  
compatible with free boundary conditions
should be modified if the Poisson structure is noncanonical.
The new criterion requires cancellation  of 
boundary contributions
to the Hamiltonian equations of motion. In the same time, 
boundary contributions
to the variation of Hamiltonian are allowed. The Ashtekar formalism
for gravity and hydrodynamics of the ideal fluid with a free
surface in the Clebsch variables are treated as examples.


\end{abstract}
\end{English}
\begin{Russian}
\title{О различии между допустимыми
и ``дифференцируемыми'' гамильтонианами}
\author{В.О.~Соловьев\\
\small Институт физики высоких энергий\\
\small  Протвино, 142284, Московская область, Россия\\
\small e-mail: vosoloviev@mx.ihep.su}
\date{30 октября 1996}
\maketitle
\begin{abstract}
Показано, что критерий Редже-Тейтельбойма, однозначно фиксирующий
поверхностные члены в гамильтониане, так чтобы они были совместимы
со свободными граничными условиями, нуждается в пересмотре в случае
неканонической пуассоновой структуры. Новый критерий требует
компенсации граничного вклада в гамильтоновы уравнения движения.
В то же время, граничные вклады в вариацию гамильтониана допускаются.
В качестве примеров рассмотрены формализм Аштекара для теории гравитации
и гидродинамика идеальной жидкости со свободной границей в переменных
Клебша. 
\end{abstract}
\end{Russian}
\newpage
\begin{English}
In this publication 
we consider the case of free boundary conditions in field theory, i.e., the
situation when variations of the field variables and their spatial
derivatives are not necessarily equal to zero on the boundary of the relevant
space domain. Therefore 
it would be incorrect to treat Hamiltonians (or Lagrangians)
differing in spatial divergences as representatives of the same equivalence
class. We are to fix a unique
admissible Hamiltonian (or Lagrangian) 
from that class according to some rule.
\end{English}
\begin{Russian}
В этой работе мы рассматриваем случай свободных граничных условий, то есть,
ситуацию, когда вариации полевых переменных и их пространственные производные
могут быть отличны от нуля на границе рассматриваемой области пространства.
Тогда было бы ошибочным трактовать гамильтонианы (или лагранжианы),
отличающиеся на пространственную дивергенцию, как принадлежащие одному
классу эквивалентности. Необходимо зафиксировать единственный допустимый
гамильтониан (или лагранжиан) из этого класса в соответствии с некоторым
правилом.
\end{Russian}

\begin{English}
The importance of such problems for physics can be easily seen from
the long history of discussion on the role of surface integrals in
the Hamiltonian of General Relativity
\end{English}
\begin{Russian}
Важность подобных задач для физики легко понять, например, из 
истории длительной дискуссии о роли поверхностных интегралов в
гамильтониане общей теории относительности
\end{Russian}
\cite{ADM} --
\cite{DeWitt}. 
\begin{English}
This discussion had been resolved in a publication
by Regge and Teitelboim
\end{English}
\begin{Russian}
Точка в этой дискуссии была поставлена работой Редже и Тейтельбойма,
\end{Russian}
\cite{RT}
\begin{English}
where a criterion for the adequate choice of divergences
in the Hamiltonian  was proposed. It was called the requirement of
``differentiability'' of the Hamiltonian (see also
\end{English}
\begin{Russian}
где был предложен критерий адекватного выбора дивергенций в гамильтониане.
Он получил название требования ``дифференцируемости'' гамильтониана
(см. также
\end{Russian}
\cite{FT}
).

\begin{English}
Here we intend to show that this criterion, which is applicable for
the canonical Poisson structure,  should be revised in a more general
situation, especially if boundary contributions appear in the symplectic
form and (or) in the Poisson brackets.
In the same time the general mathematical background
staying behind the concrete formulation given to it by Regge and Teitelboim
is untouched by our improvement.
\end{English}
\begin{Russian}
Мы собираемся показать здесь, что этот критерий, справедливый
при канонической
пуассоновой структуре, должен быть пересмотрен в более общей ситуации,
особенно когда граничные вклады появляются в симплектической форме и (или)
в скобках Пуассона. В то же время общая математическая основа, стоящая за
конкретной формулировкой, данной ей Редже и Тейтельбоймом, остается 
неизменной и в нашем подходе.
\end{Russian}

\begin{English}
This general background is the ``natural boundary conditions'' criterion
of the variational calculus
\end{English}
\begin{Russian}
Этой общей основой является критерий ``естественных граничных условий''
в вариационном исчислении
\end{Russian}
\cite{HC}, \cite{Lancz}.
\begin{English}
The general idea of the
natural boundary conditions is
that we can get from a variational principle not only the Euler-Lagrange
equations
but also some equations on the boundary.
Both of them follow from the requirement
that the functional under study should have a stationary point.
It is possible to derive the natural boundary conditions 
when arbitrary field variations on the boundary contribute to the
functional variation,
and so, their coefficients must be equal to
zero to make the functional stationary.
\end{English}
\begin{Russian}
Основная идея естественных граничных условий состоит в том, что из
вариационного принципа можно получить не только уравнения Эйлера-Лагранжа,
но и некоторые граничные уравнения. И те и другие следуют из требования,
чтобы изучаемый функционал имел стационарную точку. Можно вывести
естественные граничные условия, когда произвольные вариации полей 
на границе дают вклад в вариацию функционала,  таким образом, 
их коэффициенты должны обращаться в ноль,
чтобы обеспечить его стационарность. 
\end{Russian}

\begin{English}
The difference between admissible and ``differentiable'' Hamiltonians,
which is discussed in the paper, arises when noncanonical variables
are used in the Hamiltonian formalism. This situation is not unfamiliar.
For example, noncanonical Poisson brackets may originate as Dirac
brackets after some reduction procedure. Below we will consider two
examples. The first one is the Ashtekar formulation of General Relativity
where the noncanonicity of Poisson brackets 
arises as a result of the change of variables
\end{English}
\begin{Russian}
Различие между допустимыми и ``дифференцируемыми'' 
гамильтонианами, обсуждаемое
в настоящей работе, возникает тогда, когда в гамильтоновом формализме
появляются неканонические переменные. Эта ситуация не является необычной.
Например, неканонические скобки Пу\-ассона могут возникнуть как скобки Дирака
после проведения процедуры редукции. Ниже будет рассмотрено два примера.
Первым из них служит формализм Аштекара в теории гравитации, где
неканонические скобки Пуассона появляются после преобразования переменных
\end{Russian}
\cite{Sol92}.
\begin{English}
The second example is the Hamiltonian formalism for surface waves in
ideal fluid. Here just a position of the boundary becomes a dynamical
variable. In both cases we will see that the relation between
the Lagrangian and the Hamiltonian is not so simple as in the 
canonical situation.
As a result we can see that the correspondence between the action functional
and the boundary conditions 
survives in the general case whereas the requirement
of ``differentiability'' for the Hamiltonian should be replaced by another one.
The new criterion requires a cancellation of 
boundary terms in the
Hamiltonian vector field, or saying it in other words, in the Hamiltonian
equations of motion. The Hamiltonian vector field is to be constructed
according to the so-called formal variational calculus
\end{English}
\begin{Russian}
Вторым примером является гамильтонов формализм для поверхностных волн
в идеальной жидкости, где само положение границы становится динамической
переменной. В обоих случаях мы увидим, что связь между лагранжианом и
гамильтонианом не так очевидна, как это имеет место в канонических
переменных. В результате будет показано, что соответствие между функционалом
действия и граничными условиями сохраняется в общем случае, тогда как
требование ``дифференцируемости'' гамильтониана должно быть заменено
другим. Новый критерий требует сокращения граничных членов в гамильтоновом
векторном поле, или, иначе говоря, в гамильтоновых уравнениях движения.
Гамильтоново векторное поле должно строиться согласно правилам так 
называемого формального вариационного исчисления
\end{Russian}
\cite{GD}
\begin{English}
and its extension on divergences proposed in previous publications
of the author 
\end{English}
\begin{Russian}
и его расширения на дивергенции, предложенного в предыдущих публикациях
автора
\end{Russian}
\cite{Sol93}, \cite{Sol95}, \cite{Sol96}.

\begin{English}
Throughout this paper we will use the characteristic function of a
relevant compact space domain 
\end{English}
\begin{Russian}
Ниже всюду мы будем использовать характеристическую функцию для
соответствующей компактной пространственной области
\end{Russian}
$\Omega$
\begin{English}
constructed by means of the Heaviside 
\end{English}
\begin{Russian}
построенную с помощью 
\end{Russian}
$\theta$
\begin{English}
-function
and smooth function 
\end{English}
\begin{Russian}
-функции Хевисайда и гладкой функции
\end{Russian}
$P(x)$ 
\begin{English}
with the following properties
\be
P(x)= \cases {
>0 &  if $x\in\Omega$; \cr
=0  & if $x\in\partial\Omega$; \cr
<0 & otherwise. \cr }
\ee
\end{English}
\begin{Russian}
со следующими свойствами
\be
P(x)= \cases {
>0 &  если $x\in\Omega$; \cr
=0  & если $x\in\partial\Omega$; \cr
<0 & иначе. \cr }
\ee
\end{Russian}
\begin{English}
It allows us to write any integral over 
\end{English}
\begin{Russian}
Это позволяет записывать любой интеграл по области
\end{Russian}
$\Omega$ 
\begin{English}
formally as an integral over infinite space 
\end{English}
\begin{Russian}
формально в виде интеграла по всему бесконечному пространству
\end{Russian}
${\rm R}^3$ 
\begin{English}
and freely integrate by parts. We will omit
\end{English}
\begin{Russian}
и свободно интегрировать по частям. Мы будем опускать
\end{Russian}
$d^3x$ 
\begin{English}
in these integrals.
\end{English}
\begin{Russian}
в этих интегралах.
\end{Russian}

\begin{English}
As a first example let us consider the Ashtekar formalism for canonical
gravity
\end{English}
\begin{Russian}
В качестве первого примера рассмотрим канонический формализм Аштекара для
гравитации
\end{Russian}
\cite{Ash}. 
\begin{English}
It is well known that this formulation can be constructed as a
result of field transformations starting from the tetrad variables in time 
gauge where the unit normal to spacelike hypersurface is taken as one leg
of the tetrad
\end{English}
\begin{Russian}
Хорошо известно, что этот формализм может быть построен в результате
преобразований полевых переменных, начиная с тетрад во временн\'ой калибровке,
когда единичная нормаль к пространственно-подобной гиперповерхности
принимается в качестве одного из векторов тетрады
\end{Russian}
\cite{Henn}.
\begin{English}
The initial pair of canonical variables 
\end{English}
\begin{Russian}
Исходная пара канонических переменных
\end{Russian}
$(E_{ia},\pi^{ia})$
\begin{English}
can first be changed for another one 
\end{English}
\begin{Russian}
может быть сначала заменена другой парой
\end{Russian}
$(\tilde E^{ia},K_{ia})$
\be
\{\tilde E^{ia}(x),K_j^b(y)\}=\frac{1}{2}\delta^i_j\delta^{ab}\delta(x,y),
\ee
\begin{English}
where
\end{English}
\begin{Russian}
где
\end{Russian}
\be
\tilde E^{ia}=EE^{ia},\quad K^a_i=K_{ij}E^{ja}+E^{-1}E_{ib}J^{ab},
\ee
\begin{English}
and 
\end{English}
\begin{Russian}
и
\end{Russian}
$E^a_iE^j_a=\delta^j_i$,
$E^a_iE^i_b=\delta^a_b$, $E=\det|E_{ia}|$, $K_{ij}$ 
\begin{English}
is the second fundamental
form of the spacelike hypersurface,
and the generator of triad (the three remaining vectors of the tetrad)
rotations is
\end{English}
\begin{Russian}
является второй фундаментальной формой пространственно-подобной
гиперповерхности, а генератор вращений триады (то есть, тройки остальных
векторов тетрады) есть
\end{Russian}
\be
J^{ab}=\frac{1}{2}(K^a_i\tilde E^{bi}-K^b_i\tilde E^{ai}).\label{eq:Jab}
\ee

\begin{English}
Then the second transformation is used which introduce the Ashtekar
connection variable 
\end{English}
\begin{Russian}
Затем используется второе преобразование, которое вводит переменную
связности Аштекара
\end{Russian}
$A^a_i$ 
\begin{English}
instead of 
\end{English}
\begin{Russian}
вместо
\end{Russian}
$K_{ia}$
\be
A^a_i=iK^a_i+\Gamma^a_i,\qquad
\Gamma^a_i=\frac{1}{2}\epsilon^{abc}\tilde E_{jc}\tilde E^{jb}_{\ |i},
\ee
\begin{English}
where the vertical line denotes the standard Riemannian covariant
derivative and
\end{English}
\begin{Russian}
где вертикальной линией обозначена стандартная риманова ковариантная 
производная и
\end{Russian}
\be
\{\tilde
E^{ia}(x),A^b_j(y)\}=\frac{i}{2}\delta^i_j\delta^{ab}\delta(x,y).
\ee

\begin{English}
This transformation leads to the new form of the internal rotation
generator which we take as a primer of the admissible Hamiltonian
\end{English}
\begin{Russian}
Это преобразование приводит к другой записи генератора внутренних
вращений, который мы выбираем в качестве примера допустимого гамильтониана
\end{Russian}
\be
H(\hat\lambda^a)=
2\int \theta(P)\hat\lambda^a{\cal D}_i\tilde E^{ia}\equiv 2i
\int \theta(P)\hat\lambda^a\epsilon^{abc}J^{bc}.
\ee
\begin{English}
Here the covariant derivative 
\end{English}
\begin{Russian}
Ковариантная производная
\end{Russian}
${\cal D}_i$ 
\begin{English}
is defined by the following relation
\end{English}
\begin{Russian}
определяется следующим соотношением
\end{Russian}
\be
{\cal D}_i\xi^{ka}=\xi^{ka}_{\ |i}+\epsilon^{abc}A^b_i\xi^{kc}.
\ee

\begin{English}
In contrast to its previous form given according to the formula 
\end{English}
\begin{Russian}
В отличие от его прежнего вида, заданного формулой
\end{Russian}
(\ref{eq:Jab}), 
\begin{English}
in the Ashtekar variables the generator density depends on
the spatial derivatives of 
\end{English}
\begin{Russian}
в переменных Аштекара плотность генератора зависит от пространственных
производных величины
\end{Russian}
$\tilde E^{ia}$, 
\begin{English}
and so its variation 
\end{English}
\begin{Russian}
и поэтому его вариация
\end{Russian}
\be
\delta H=
\int \left(\frac{\delta H}{\delta\tilde E^{ia}}\delta\tilde E^{ia}
+\frac{\delta H}{\delta A^a_i}\delta A^a_i
\right),
\ee
\begin{English}
contains a boundary contribution
\end{English}
\begin{Russian}
содержит граничный вклад
\end{Russian}
\begin{eqnarray}
\frac{\delta H}{\delta\tilde E^{ia}}&=&-\theta_{,i}2\hat\lambda^a-
\theta 2{\cal D}_i\hat\lambda^a,\label{eq:eder}\\
\frac{\delta H}{\delta A^a_i}&=&-\theta 2\epsilon^{abc}\hat\lambda^b
\tilde E^{ic}.
\end{eqnarray}
\begin{English}
If we deal with a free boundary condition case then this functional is
not  ``differentiable'' in the Regge-Teitelboim terminology.
Now we will show that it is nevertheless admissible as it
gives regular Hamiltonian equations. 

The explanation follows from the fact that the Ashtekar variables are
canonical only up to the boundary term
\end{English}
\begin{Russian}
При свободных граничных условиях этот функционал не является
``дифференцируемым'' в терминологии Редже-Тейтельбойма. Мы сейчас
покажем, что тем не менее он допустим, так как приводит к регулярным
гамильтоновым уравнениям.

Объясняется это тем, что переменные Аштекара являются
каноническими лишь с точностью до граничного члена
\end{Russian}
\cite{Sol92}
\be
\left\{A^a_i(x),A^b_j(y)\right\}=\theta_{,k}C^k_{[(ia)(jb)]}\delta(x,y),
\ee
\begin{English}
where the square brackets denote antisymmetrization 
\end{English}
\begin{Russian}
где квадратные скобки обозначают антисимметризацию
\end{Russian}
$i\leftrightarrow j$,
$a\leftrightarrow b$ 
\begin{English}
and
\end{English}
\begin{Russian}
и
\end{Russian}
\be
C^k_{(ia)(jb)}=\frac{i}{2E}(\epsilon^{acb}\delta^k_jE_{ic}-
\epsilon^{acd}E_{ib}E_{jc}E^{kd}).
\ee

\begin{English}
Due to this noncanonicity the Hamiltonian equations aquire a form
\end{English}
\begin{Russian}
Благодаря этой неканоничности гамильтоновы уравнения записываются
в виде
\end{Russian}
\begin{eqnarray}
\dot {\tilde E}^{ia}(x)&=&\int \{\tilde E^{ia}(x),A^b_j(y)\}
\frac{\delta H}{\delta
A^b_j(y)},\\
\dot A^a_i(x)&=&\int \{A^a_i(x),\tilde E^{jb}(y)\}
\frac{\delta H}{\delta\tilde E^{jb}(y)}+\int\{A^a_i(x),A^b_j(y)\}
\frac{\delta H}{\delta A^b_j(y)},
\end{eqnarray}
\begin{English}
or, in the explicit form obtained by using the prescription from
papers
\end{English}
\begin{Russian}
или в явном виде, полученном с использованием рецептов из работ
\end{Russian}
\cite{Sol93}-\cite{Sol96}, 
\begin{English}
(here it is simply 
\end{English}
\begin{Russian}
(здесь это есть просто соотношение
\end{Russian}
$\theta(P)\cdot\theta'(P)=\theta'(P)$), 
\begin{English}
they are
\end{English}
\begin{Russian}
они принимают вид
\end{Russian}
\begin{eqnarray}
\dot{\tilde E}^{ia}&=& i\epsilon^{abc}\hat\lambda^c\tilde E^{ib},\\
\dot A^a_i&=& i{\cal D}_i\hat\lambda^a,
\end{eqnarray}
\begin{English}
where 
\end{English}
\begin{Russian}
где
\end{Russian}
$\theta (P)$
\begin{English}
-factors are omitted.

We can see that the singular on the boundary
terms in the second equation have been mutually
canceled despite their presence in the full variational derivative 
\end{English}
\begin{Russian}
-факторы опущены при записи.

Отсюда видно, что сингулярные на границе члены во втором уравнении
взаимно уничтожаются, несмотря на их присутствие в полной вариационной
производной
\end{Russian}
(\ref{eq:eder}). 
\begin{English}
It means that our Hamiltonian is admissible under
arbitrary boundary conditions.

It is not surprising then that the Poisson algebra of these generators is
closed irrespectable to boundary conditions
\end{English}
\begin{Russian}
Это означает, что наш гамильтониан является допустимым при любых граничных
условиях.

Неудивительно поэтому, что алгебра скобок Пуассона для этих генераторов
замыкается независимо от граничных условий
\end{Russian}
\begin{equation}
\{H(\hat\xi^a),H(\hat\eta^b)\}=
H\left(i\epsilon^{cab}\hat\xi^a\hat\eta^b\right).
\end{equation}

\begin{English}
In the Lagrangian treatment it is evident that due to the absence
of spatial derivatives in the action
\end{English}
\begin{Russian}
В лагранжевом подходе очевидно, что поскольку действие
\end{Russian}
\begin{equation}
S=2\int\limits_{t_1}^{t_2}dt\int\theta(P)\left(K^a_i\dot{\tilde E^{ia}}-
i\hat\lambda^a\varepsilon^{abc}J^{bc}\right),
\end{equation}
\begin{English}
any boundary conditions are natural. The Ashtekar transformation
gives it a new appearance
\end{English}
\begin{Russian}
вовсе не содержит пространственных производных, любые граничные
условия являются естественными. Замена переменных, предложенная
Аштекаром, представляет это действие в виде
\end{Russian}
\begin{equation}
S=-2i\int\limits_{t_1}^{t_2}dt\int\theta(P)\left((A^a_i-\Gamma^a_i)
\dot{\tilde E^{ia}}-
i\hat\lambda^a{\cal D}_i\tilde E^{ia}\right),
\end{equation}
\begin{English}
where this property is no more evident. But let us derive a
variation of the action
\end{English}
\begin{Russian}
из которого указанное свойство уже не очевидно. Найдем, однако, вариацию
\end{Russian}
\begin{eqnarray}
\delta S&=&-2i\int\limits_{t_1}^{t_2}dt\int\theta(P)
\Biggl(\delta{\tilde E^{ia}}
\left[-\dot A^a_i+i{\cal D}_i\tilde E^{ia}\right]+\delta A^a_i\left[
\dot{\tilde E^{ia}}+i\varepsilon^{abc}\lambda^b\tilde E^{ic}\right]+\nonumber\\
&+&\frac{\partial}{\partial t}\left((A^a_i-\Gamma^a_i)
\delta\tilde E^{ia}\right)+\left((\dot\Gamma^a_i\delta\tilde E^{ia}-
\dot{\tilde E^{ia}}\delta\Gamma^a_i)-\partial_i(i\hat\lambda^a
\delta\tilde E^{ia})\right)\Biggr).
\end{eqnarray}
\begin{English}
Then the total time derivative term gives zero contribution as field
variations on the time interval boundary are zero, the terms in square brackets
give the equations of motion and the rest terms give no contribution
due to relation
\end{English}
\begin{Russian}
Тогда член с полной производной по времени дает нулевой вклад поскольку
на границах временного интервала вариации координат обращаются в ноль,
члены в квадратных скобках дают уравнения движения, а остальные члены
не дают вклада, поскольку
\end{Russian}
\begin{equation}
(\dot\Gamma^a_i\delta\tilde E^{ia}-
\dot{\tilde E^{ia}}\delta\Gamma^a_i)-\partial_i(i\hat\lambda^a
\delta\tilde E^{ia})=0\ {\rm mod}\left(\dot{\tilde E^{ia}}
+i\varepsilon^{abc}\lambda^b\tilde E^{ic}=0\right),
\end{equation}
\begin{English}
valid for arbitrary functions
\end{English}
\begin{Russian}
для любых функций
\end{Russian}
$\delta\tilde  E^{ia}$,
\begin{English}
that can be verified by the straightforward calculation.
\end{English}
\begin{Russian}
что проверяется прямым вычислением.
\end{Russian}

\begin{English}
As the second example we consider the Hamiltonian description of ideal
fluid with a free surface. In Eulerian variables the action can be
written
\end{English}
\begin{Russian}
В качестве второго примера рассмотрим гамильтоново описание идеальной
жидкости со свободной поверхностью. В эйлеровых переменных действие
может быть записано 
\end{Russian}
\cite{SW}
\begin{English}
by means of Clebsch potentials
\end{English}
\begin{Russian}
при помощи потенциалов Клебша
\end{Russian}
\be
{\bf v}=\nabla\phi+\frac{\eta}{\rho}\nabla s+\frac{\beta}{\rho}\nabla\alpha,
\ee
\begin{English}
in the following way
\end{English}
\begin{Russian}
следующим образом
\end{Russian}
\begin{equation}
S=\int\limits_{t_1}^{t_2}dt\int\theta(P)\left[
\rho\left(\frac{{\bf v}^2}{2}-\Phi({\bf x})-
\varepsilon(\rho,s)\right)-\rho\frac{D\phi}{Dt}
-\eta\frac{Ds}{Dt}-
\beta\frac{D\alpha}{Dt}-\tau K\right].\label{eq:action_Efbtrue}
\end{equation}
\begin{English}
where 
\end{English}
\begin{Russian}
где
\end{Russian}
$\rho$ 
\begin{English}
is the fluid mass density, 
\end{English}
\begin{Russian}
- плотность массы жидкости,
\end{Russian}
$s$ 
\begin{English}
is the specific entropy,
\end{English}
\begin{Russian}
-- энтропия на единицу массы,
\end{Russian}
$\varepsilon=\varepsilon(\rho,s)$ 
\begin{English}
is the specific internal energy density,
\end{English}
\begin{Russian}
-- плотность внутренней энергии,
\end{Russian}
$\tau$ 
\begin{English}
is the surface tension coefficient,
\end{English}
\begin{Russian}
-- коэффициент поверхностного натяжения,
\end{Russian}
\be
\frac{D}{Dt}=\frac{\partial}{\partial t}+{\bf v}\cdot\nabla,\qquad
K=-\nabla\cdot\left(\frac{\nabla P}{|\nabla P|}\right).
\ee
\begin{English}
On the boundary the last formula gives the external curvature of it
whereas 
\end{English}
\begin{Russian}
На границе последняя формула дает ее внешнюю кривизну, тогда как
\end{Russian}
$\nabla P$ 
\begin{English}
is proportional to the normal, the proposal to use it
is due to Abarbanel et al
\end{English}
\begin{Russian}
пропорционален вектору нормали, предложение использовать эти формулы
исходит от Абарбанеля и др.
\end{Russian}
\cite{Abarb}.

\begin{English}
The corresponding symplectic form 
\end{English}
\begin{Russian}
Соответствующая симплектическая форма
\end{Russian}
\begin{equation}
-\int \left[ \theta(P)(\delta\rho\wedge\delta\phi+
\delta\eta\wedge\delta s+\delta\beta\wedge\delta\alpha)+
\theta'(P)\delta P\wedge(\rho\delta\phi+\eta\delta s+\beta
\delta\alpha)\right].\label{eq:symplectic}
\end{equation}
\begin{English}
is degenerate. To go to the Hamiltonian formalism we can use the Dirac
procedure
\end{English}
\begin{Russian}
является вырожденной. Для перехода к гамильтонову формализму можно
воспользоваться процедурой Дирака
\end{Russian}
\cite{Dirac64} 
\begin{English}
or the Faddeev-Jackiw approach
\end{English}
\begin{Russian}
или подходом Фаддеева-Джакива
\end{Russian}
\cite{FJ},
\begin{English}
but really both of them lead to the same result as a simple trick.
If we introduce  a canonical variable 
\end{English}
\begin{Russian}
но в действительности оба пути ведут к тому же результату, что и
простейший трюк. Если ввести каноническую переменную
\end{Russian}
$\pi$, 
\begin{English}
conjugate to 
\end{English}
\begin{Russian}
сопряженную к
\end{Russian}
$P$
\be
S\rightarrow S+\int\limits^{t_2}_{t_1}dt\int (\pi\dot P-
\lambda\pi),
\ee
\begin{English}
and so add to the symplectic form the standard term
\end{English}
\begin{Russian}
и таким образом дополнить симплектическую форму стандартным членом
\end{Russian}
\be
\int \delta\pi\wedge \delta P,
\ee
\begin{English}
then the modified symplectic form becomes nondegenerate and can be
inverted for getting the Poisson bivector
\end{English}
\begin{Russian}
то в модифицированном виде симплектическая форма становится невырожденной
и может быть обращена, что позволяет получить выражение для пуассонова
бивектора
\end{Russian}
\begin{eqnarray}
\Psi&=&\int \Biggl[\theta(P)\left(
\frac{\delta}{\delta\rho}\wedge\frac{\delta}{\delta\phi}+
\frac{\delta}{\delta\eta}\wedge\frac{\delta}{\delta s}+\frac{\delta}
{\delta\beta}\wedge\frac{\delta}{\delta\alpha}
\right)+\nonumber\\
&+&\frac{\delta}{\delta P}\wedge\frac{\delta}{\delta\pi}+
\theta'(P)\frac{\delta}{\delta\pi}\wedge\left(
\rho\frac{\delta}{\delta\rho}+\eta\frac{\delta}{\delta\eta}+
\beta\frac{\delta}{\delta\beta}\right)\Biggr],
\end{eqnarray}
\begin{English}
which explicitely contains the boundary 
\end{English}
\begin{Russian}
который явным образом содержит граничную
\end{Russian}
$\delta$
\begin{English}
-function.
\end{English}
\begin{Russian}
-функцию.
\end{Russian}

\begin{English}
The Hamiltonian has the following form
\end{English}
\begin{Russian}
Гамильтониан имеет следующий вид
\end{Russian}
\begin{equation}
{\rm H}=\int \left[\theta(P)\left(\frac{\rho{\bf v}^2}{2}
+\rho\Phi+\rho\varepsilon
(\rho,s)+\tau K\right)+\lambda \pi\right],
\end{equation}
\begin{English}
and its variation also have singular boundary contributions,
for example,
\end{English}
\begin{Russian}
и его вариация также содержит сингулярные граничные вклады, например,
\end{Russian}
\begin{eqnarray}
\frac{\delta{\rm H}}{\delta P}&=&\theta'(P)\left(
\frac{\rho {\bf v}^2}{2}+\rho\Phi+\rho\varepsilon+\tau K
\right),\\
\frac{\delta{\rm H}}{\delta\phi}&=&-\theta'(P)\rho{\bf v}\cdot\nabla P-
\theta(P)\nabla(\rho{\bf v}),
\end{eqnarray}
\begin{English}
and so on.
\end{English}
\begin{Russian}
и так далее.
\end{Russian}

\begin{English}
Let us estimate the Hamiltonian vector field according to
the standard formula where the interior product should be
understood according to the definition
given in Refs.
\end{English}
\begin{Russian}
Найдем теперь гамильтоново векторное поле согласно стандартной формуле,
где внутреннее умножение должно пониматься в смысле определения, данного
в работе
\end{Russian}
\cite{Sol95}, \cite{Sol96}
\begin{eqnarray}
-\d{\rm H}\inprod\Psi
&=&
\left[-\theta'(P)\rho(\lambda+{\bf v}\cdot\nabla P)-\theta(P)
\nabla(\rho{\bf v})\right]\frac{\delta}{\delta\rho}+
\nonumber\\
&+&\left[-\theta'(P)\eta(\lambda+{\bf v}\cdot\nabla P)-
\theta(P)(\nabla(\eta{\bf v})-\rho T)\right]\frac{\delta}{\delta\eta}
+\nonumber\\
&+&\left[-\theta'(P)\beta(\lambda+{\bf v}\cdot\nabla P)-\theta(P)
\nabla(\beta{\bf v})\right]\frac{\delta}{\delta\beta}
+\nonumber\\
&+&\theta'(P)(p-\tau K)\frac{\delta}
{\delta\pi}
+\nonumber\\
&+&\theta(P)\left[\frac{v^2}{2}-{\bf v}\cdot\nabla\phi-\Phi-
\varepsilon-\frac{p}{\rho}\right]\frac{\delta}{\delta\phi}
-\theta(P)\left[{\bf v}\cdot\nabla s\right]\frac{\delta}{\delta s}
-\nonumber\\
&-&\theta(P){\bf v}\cdot\nabla\alpha\frac{\delta}{\delta\alpha}
+\theta(P)\lambda\frac{\delta}{\delta P},
\end{eqnarray}
\begin{English}
here 
\end{English}
\begin{Russian}
здесь
\end{Russian}
$p=\rho^2\frac{\partial\varepsilon}{\partial\rho}$ 
\begin{English}
is the pressure and 
\end{English}
\begin{Russian}
-- давление и
\end{Russian}
$T=\frac{\partial\varepsilon}{\partial s}$ 
\begin{English}
is the temperature.

Then the requirement that this Hamiltonian vector field must  not contain
boundary terms is equivalent to the standard boundary conditions
for the problem
\end{English}
\begin{Russian}
-- температура.

Тогда требование того, что гамильтоново векторное поле не должно содержать
граничных членов, эквивалентно стандартным для этой задачи граничным
условиям 
\end{Russian}
\cite{Lamb}
\begin{eqnarray}
\theta'(P)(\dot P+{\bf v}\cdot\nabla P)&=&0,\label{eq:1}\\
\theta'(P)(p-\tau K)&=&0,\label{eq:2}
\end{eqnarray}
\begin{English}
if we take into account equation of motion
\end{English}
\begin{Russian}
если мы примем во внимание уравнение движения
\end{Russian}
\begin{equation}
\dot P=
\lambda.
\end{equation}
\begin{English}
It is useful to compare this approach with the analisys made in other
variables
\end{English}
\begin{Russian}
Полезно сравнить данный подход с анализом, проведенным на основе других
переменных
\end{Russian}
\cite{Abarb}.

\begin{English}
If we try to use the criterion by Regge and Teitelboim here we will get the
wrong boundary conditions, for example, one of them will be
\end{English}
\begin{Russian}
Если бы мы попытались использовать здесь критерий Редже и Тейтельбойма,
то пришли бы к неверным граничным условиям, например, одно из них
имело бы вид
\end{Russian}
\be
{\bf v}\cdot\nabla P=0,
\ee
\begin{English}
that is, the requirement for the boundary to be fixed.
\end{English}
\begin{Russian}
то есть, это было бы условие неподвижности границы.
\end{Russian}

\begin{English}
Lagrangian approach to the problem consists in studying the variation
of the action (\ref{eq:action_Efbtrue}). Apart from terms giving equations
of motion this variation also contains boundary terms
\end{English}
\begin{Russian}
В лагранжевом подходе к обсуждаемой проблеме изучается вариация действия
(\ref{eq:action_Efbtrue}). Помимо членов, дающих стандартные уравнения
движения, эта вариация также содержит граничные члены
\end{Russian}
\begin{eqnarray}
\delta'S=\int\limits^{t_2}_{t_1}dt\int\Biggl(-\frac{\partial}{\partial t}\Bigl(
\theta\rho\delta\phi+\theta\eta\delta s+\theta\beta\delta\alpha\Bigr)
+\nonumber\\
+\theta'\left(\dot P+{\bf v}\cdot\nabla P\right)\left(\rho\delta\phi+
\eta\delta s+\beta\delta\alpha\right)+\theta'(p-\tau K)\delta P\Biggr).
\end{eqnarray}
\begin{English}
The total time derivative gives the symplectic form (\ref{eq:symplectic})
and does not contribute to the variational principle because the field
variations are zero on the time boundary. The other terms contribute
at the spatial boundary where the field variations are arbitrary
and just give us the natural boundary conditions which are the same as 
(\ref{eq:1}), (\ref{eq:2}).
\end{English}
\begin{Russian}
Полная производная по времени позволяет найти симплектическую форму
(\ref{eq:symplectic}), но не дает вклада в вариационный принцип, так как
вариации полей равны нулю на временн\'ой границе. Оставшиеся члены
существенны на пространственной границе, где вариации полей являются
произвольными, как раз эти члены и позволяют найти 
естественные граничные условия, совпадающие с
(\ref{eq:1}), (\ref{eq:2}).
\end{Russian}

\begin{English}
We have shown that in general situation when the Hamiltonian variables
are not necessarily canonical and their Poisson brackets may contain
boundary terms the Regge-Teitelboim criterion of
``differentiability'' of the Hamiltonian must be replaced by a new one.
A Hamiltonian may be considered as admissible if the Hamiltonian vector
field constructed 
according to the extended definitions given in 
\end{English}
\begin{Russian}
Мы показали, что в общем случае, когда гамильтоновы переменные не обязательно
являются каноническими и их скобки Пуассона могут содержать граничные члены,
критерий ``дифференцируемости'' гамильтониана, предложенный Редже и 
Тейтельбоймом, должен быть заменен новым критерием. Гамильтониан может
считаться допустимым, если гамильтоново векторное поле, построенное
согласно обобщенным определениям, данным в работах
\end{Russian}
\cite{Sol95}, \cite{Sol96}, 
\begin{English}
does not
contain any boundary contribution. This completes the search for a new
criterion started in 
\end{English}
\begin{Russian}
не содержит никаких граничных вкладов. Этим решается задача поиска нового
критерия, поставленная в работе
\end{Russian}
\cite{Sol92}. 
\begin{English}
In general form the idea can be also
traced to a publication by Mason 
\end{English}
\begin{Russian}
В общей форме идея такого решения может быть прослежена также в работе
Мейсона
\end{Russian}
\cite{Mason}.
\begin{English}
More detailed treatment will be given elsewhere.

We hope that the Hamiltonian approach to field theory with free
boundary conditions will be useful in dealing with different physical
problems, especially those where the Lagrangian approach meets with
difficulties.
\end{English}
\begin{Russian}
Более подробное обсуждение будет дано в другом месте.

Мы надеемся, что гамильтонов подход к задачам теории поля со свободными
граничными условиями окажется полезен при рассмотрении различных физических
проблем,  особенно тех, где лагранжев подход встречается с затруднениями.
\end{Russian}

\vspace{12pt}
\begin{English}
{\large\bf Acknowledgements}

This work has been completed during a visit of the author to the International
Centre for Theoretical Physics in Trieste. The author is most grateful
to Professor S.~Randjbar-Daemi for the invitation and kind hospitality, 
partial support from ICTP is gratefully acknowledged. It is a pleasure
to thank A.~Zheltukhin for discussion.
\end{English}
\begin{Russian}
{\large\bf Благодарности}

Эта работа была закончена во время визита автора в международный центр
теоретической физики в Триесте. Автор благодарен проф. С.~Ранджбару-Даеми
за приглашение и гостеприимность. Поддержка со стороны МЦТФ признается
с благодарностью. Приятно поблагодарить проф. А.~Желтухина за обсуждение
данной работы.
\end{Russian}

\newpage
\hfill


\begin{thebibliography}{**}

\bibitem{ADM}
R.~Arnowitt, S.~Deser and Ch.W.~Misner, in: Gravitation, an Introduction
to Current Research, ed. L.~Witten (New York, 1963).
\begin{Russian}
(Имеется русский перевод: Эйнштейновский сборник 1967, Наука,
М., 1967)
\end{Russian}

\bibitem{Dirac}
P.A.M.~Dirac, { Phys. Rev. Lett.} { 2} (1959) 368;
{ Phys. Rev.} { 114} (1959) 924.

\bibitem{Higgs}
P.W.~Higgs, { Phys. Rev. Lett.} { 3} (1959) 66.

\bibitem{Schwinger}
J.~Schwinger, { Phys. Rev.} { 130} (1963) 1253.
\begin{Russian}
(Имеется русский перевод: Гравитация и топология. Актуальные проблемы.
Сборник статей под ред. Д.Д.~Иваненко, Мир, М., 1966)
\end{Russian}

\bibitem{DeWitt}
B.S.~DeWitt, { Phys. Rev.} { 160} (1967) 1113.

\bibitem{RT}
T.~Regge and C.~Teitelboim,
{ Ann. Phys.} { 88} (1974) 286.

\bibitem{FT}
\begin{Russian}
Л.А.~Тахтаджян, Л.Д.~Фаддеев, Гамильтонов подход в теории солитонов,
Наука, М., 1986.
\end{Russian}
\begin{English}
L.A.~Takhtajan, L.D.~Faddeev, Hamiltonian Method in Soliton Theory,
(Springer-Verlag, N.Y., 1987) pp.18-19.
\end{English}

\bibitem{HC}
R.~Courant and D.~Hilbert, Methods of Mathematical Physics,
Vol.1, (Wiley, N.Y., 1989) pp.208-211.
\begin{Russian}
(Имеется русский перевод: Д.~Гильберт, Р.~Курант, Методы
математической физики, том 1, М., 1951)
\end{Russian}

\bibitem{Lancz}
C.~Lanczos, The Variational Principles of Mechanics,
(Univ. Toronto Press, 1964) pp.68-73.
\begin{Russian}
(Имеется русский перевод: К.~Ланцош, Вариационные принципы механики,
Мир, М., 1965) 
\end{Russian}

\bibitem{Sol92}
V.O.~Soloviev, { Phys. Lett. B} { 292} (1992) 30.

\bibitem{GD}
\begin{Russian}
И.М.~Гельфанд, Л.А.~Дикий, УМН, 30 (1975) 67.
\end{Russian}
\begin{English}
I.M.~Gel'fand and L.A.~Dickey,
{ Uspekhi Matematicheskikh Nauk} { 30} (1975) 67.
\end{English}

\bibitem{Sol93}
V.O.~Soloviev, { J. Math. Phys.} { 34} (1993) 5747.

\bibitem{Sol95}
V.O.~Soloviev,
Boundary values as Hamiltonian variables II. Graded structures,
Preprint IHEP 94-145, Protvino, 1994, q-alg/9501017 (submitted
to { J. Math. Phys.}).

\bibitem{Sol96}
V.O.~Soloviev, { Nucl. Phys.} { BP49} (1996) 35.


\bibitem{Ash}
A.~Ashtekar, { Phys. Rev. Lett.}
{57} (1986) 2244;
{ Phys. Rev.} { D36}   (1987) 1587;
A.~Ashtekar, P.~Mazur, C.G.~Torre, { Phys. Rev.} { D36}
 (1987) 2955.

\bibitem{Henn}
M.~Henneaux, J.E.~Nelson, C.~Schomblond, { Phys. Rev.}
{ D39} (1989) 434.

\bibitem{SW}
R.L.~Seliger and G.B.~Whitham, { Proc. Roy. Soc.} { A305} (1968) 1.
\begin{Russian}
(Имеется русский перевод: Механика. Сборник переводов. 1969 Э5 (117) 
стр.99)
\end{Russian}

\bibitem{Abarb}
H.D.I.~Abarbanel, R.~Brown and Y.M.~Yang, { Phys. Fluids}
{ 31} (1988) 2802.

\bibitem{Dirac64}
P.~Dirac, Lectures on Quantum Mechanics, (Yeshiva Univ., N.Y., 1964).
\begin{Russian}
(Имеется русский перевод: П.А.М.~Дирак, Лекции по квантовой механике,
Мир, М., 1968)
\end{Russian}

\bibitem{FJ}
L.~Faddeev and R.~Jackiw, { Phys. Rev. Lett.} { 60} (1988) 1692.

\bibitem{Lamb}
H.~Lamb, Hydrodynamics, (Cambridge Univ. Press, Cambridge, 1932) p.456.
\begin{Russian}
(Имеется русский перевод: Г.~Ламб, Гидродинамика, ГИТТЛ, М.-Л., 1947)
\end{Russian}

\bibitem{Mason}
L.~Mason, { Class. Quant. Grav.} { 6} (1989) L7.

\end{thebibliography}
\end{document}